# Optical modes in slab waveguides with magnetoelectric effect

*Nahid Talebi*

Max-Planck Institute for Solid State Research, Stuttgart Center for Electron Microscopy, Heisenbergstr. 1, 70569 Stuttgart

E-mail: n.talebi@fkf.mpg.de

**Abstract –** Optical modes in anisotropic slab waveguides with topological and chiral magnetoelectric effects are investigated analytically, by deriving the closed-form characteristic equations of the modes and hence computing the dispersion-diagrams. In order to compute the characteristic equations, a vector-potential approach is introduced by incorporating a generalized Lorentz gauge, and the corresponding Helmholtz equations are derived correspondingly. It will be shown that the formation of the complex modes and hybridization of the optical modes in such slab waveguides is inevitable. Moreover, when the tensorial form of the permittivity in the waveguide allows for a hyperbolic dispersion, complex transition from the photonic kinds of modes to the plasmonic modes is expected.

# 1. Introduction

Electromagnetic waveguides such as optical fibres are key elements in communication networks and integrated optical circuits. The technological progress allows nowadays for the realization of highly advanced waveguides by utilizing either photonic crystal concepts [1, 2] or metamaterials [3, 4] and plasmonic [5-7]. Such waveguides are usually composed of a higher index dielectric material or metallic films embedded in semiconductors and lower index dielectric materials such as glasses. Besides the technological advances, understanding the mechanisms of waveguiding, and especially complex optical modes has been also achieved by the progress in computational and numerical techniques, such as the rigorous mode-matching technique [7], the generalized multiple technique [5, 8], the finite-difference time-domain method [6, 9], the finite-element method [10], the boundary element method [11], and the discontinuous Galerkin method [12]. However, the analytical treatments, though only possible for simpler geometries, provide us with a better understanding of the waveguiding principles, such as leaky modes [13], complex modes [14], and hybridizations [14], to only name a few.

In fact the simplest geometry which can facilitate a directional flow of the electromagnetic energy is an interface, either between a dielectric and a metal which supports surface plasmon polaritons (SPPs) [15] or between an anisotropic material and a dielectric which supports Dyakonov-Tamm waves [16]. A better control over the directionality is achieved by a slab waveguide. Despite its simplicity, the slab waveguide has been so far a prototype for understanding the mechanisms for transmission of electromagnetic energy and hybridization of optical modes. The optical modes in a symmetric isotropic slab waveguides can be decomposed into the transverse-electric (TE) and transverse magnetic (TM) waves, where each group again supports individual modes with *even* and *odd* distributions, the so-called symmetric and antisymmetric modes [17]. However, would it be possible to describe all the optical modes in an anisotropic slab waveguide by the *even* and *odd* TE and TM decompositions? What happens to these modes when the material shows the magnetoelectric (ME) effect, in either chiral [18-20] or topological [21-24] forms? These are the questions that will be addressed here, by analytically deriving the characteristic equations for the optical modes in an anisotropic slab waveguide with ME effect. In order to be able to consider all the possible modes, we directly derive the Helmholtz equation by introducing a generalized Lorentz gauge. In order to keep the discussion at a simple level, a symmetric slab waveguide is considered. Moreover, though the permittivity is allowed to be tensorial, the optic axes are taken to be aligned with the coordinate system. Besides these simplifications, it is tried to keep most of the discussions at the mathematical level and quite abstract. The reason for this is to be able to just follow the grouping of the optical modes mathematically. The formalism provided here will, however, have several applications in understanding the optical modes in anisotropic thin films such as graphite as well as chiral thin films such as $AgGaS_2$, and quartz, and moreover, to the recently discovered class of topological insulators such as Tetradymites [25], which are platforms for the realization of axion electrodynamics [24]. Discussing all these possible applications are the scope of another investigation and will be presented elsewhere.

In the following, first the necessary Helmholtz equation will be derived, which can be used to solve the optical modes in arbitrary geometrical configurations. In section 3, the optical modes at the interface of an anisotropic material with ME effect and air will be discussed. In section 4, the characteristic equations for the optical modes in anisotropic slab waveguides with ME effects will be classified. In section 5 the optical modes at the surface of a hyperbolic material with ME effect will be investigated, and finally the optical modes for an anisotropic hyperbolic slab waveguide with ME effect will be discussed in section 6.

## 2. Helmholtz equation for an anisotropic medium with ME effect

Similar to every kind of theoretical investigations of optical modes, the Maxwell equations are the starting point:

$$\vec{\nabla} \times \vec{E}(\vec{r}, \omega) = -i\omega \vec{B}(\vec{r}, \omega) \tag{1a}$$

$$\vec{\nabla} \times \vec{H}(\vec{r}, \omega) = i\omega \vec{D}(\vec{r}, \omega) \tag{1b}$$

$$\vec{\nabla} \cdot \vec{B}(\vec{r}, \omega) = 0 \tag{1c}$$

$$\vec{\nabla} \cdot \vec{D}(\vec{r}, \omega) = 0 \tag{1d}$$

where the harmonic representation as $\exp(i\omega t)$ is considered, with $i^2 = -1$. We consider also the constitutive relations including the ME effect as:

$$\vec{D}(\vec{r}, \omega) = \varepsilon_0 \hat{\varepsilon}_r(\omega) : \vec{E}(\vec{r}, \omega) + \varsigma \vec{B}(\vec{r}, \omega) \tag{2a}$$

$$\vec{H}(\vec{r}, \omega) = \frac{1}{\mu_0 \mu_r} \vec{B}(\vec{r}, \omega) - \xi \vec{E}(\vec{r}, \omega) \tag{2b}$$

where : denotes the rank two tensor multiplication. The scalars $\xi$ and $\varsigma$ appearing in the constitutive relations can be divided into Hermitian and anti-Hermitian symmetric parts:

$$\begin{aligned} \xi &= \chi - i\gamma \\ \varsigma &= \chi + i\gamma \end{aligned} \tag{3}$$

For a case of a medium showing a pure topological ME effect, $\chi = \alpha\theta/\eta_0\pi$ [21] and $\gamma = 0$, while for a pure chiral medium, $\gamma \neq 0$ and $\chi = 0$. Here $\alpha = e^2/4\pi\varepsilon_0 \hbar c$ is the fine-structure constant, $\theta$ is a phenomenological parameter in effective Ginzburg – Landau theory describing the topological magnetoelectric effect, and $\theta = \pi$ for a time – reversal invariant topological insulator with surface magnetization [21], while in general it can accept non-quantized values [26], for example for $Cr_2O_3$ [27] or for a general Tellegen medium [22]. $\eta_0 = \sqrt{\mu_0/\varepsilon_0}$ is the free-space impedance.

It has been discussed by Lekner that the constitutive relations in the form represented in Eqs. (2a) and (2b) might generate unrealistic results [28]. Instead, the constitutive relations in the form of symmetrized Condon set should be used. Interestingly, a simple transformation in the form of $\hat{\varepsilon}_r = \hat{\varepsilon}_r^{sc} - \mu_r \eta_0^2 \xi \varsigma$, $\xi = \xi^{sc}/\mu_0\mu_r$, $\varsigma = \varsigma^{sc}/\mu_0\mu_r$ reproduces the same results as discussed by Lekner. Although the symmetrized Condon set has been accepted for chiral media in many cases [29], the constitutive relations in the form of Eqs. (2a) and (2b) are directly applicable to the topological ME effect and axion electrodynamics, and has been also adopted in this paper.

Unlike the previous approaches in which the plane-wave solutions, either in the form of linearly polarized waves [30] or in the form of circularly polarized waves [19], are considered for the field components, here we construct the solutions for vector and scalar potentials $(\vec{A}, \varphi)$ in the form of Helmholtz equations. Since the magnetic flux density $\vec{B}(\vec{r}, \omega)$ is a pure solenoidal vector quantity it can be described as the curl of another vector, which is called the magnetic vector potential, as:

$$\vec{B} = \vec{\nabla} \times \vec{A} \tag{4}$$

in which we have dropped the argument $(\vec{r}, \omega)$ for simplicity. Inserting eq. (4) in eq. (1a) gives:

$$\vec{E} = -i\omega\vec{A} - \vec{\nabla}\varphi \tag{5}$$

In which $\varphi$ is the electric scalar potential. Using (1) to (5), we can derive the following equation for the magnetic vector potential:

$$\frac{1}{\mu_0 \mu_r}\left(\vec{\nabla}\vec{\nabla} \cdot \vec{A} - \nabla^2 \vec{A}\right) - \omega^2 \varepsilon_0 \hat{\varepsilon}_r : \vec{A} + \\ i\omega \varepsilon_0 \hat{\varepsilon}_r : \vec{\nabla}\varphi + i\omega(\xi - \varsigma)\vec{\nabla} \times \vec{A} = 0 \quad (6)$$

Equation (4) only defines the non-rotational part of the magnetic vector potential. We are free to employ a gauge theory to fix its solenoidal part. In order to do so, we can define a generalized Lorentz gauge in the form of:

$$\vec{\nabla}\varphi = \frac{-\hat{\varepsilon}_r^{-1}}{i\omega \varepsilon_0 \mu_0 \mu_r} : \vec{\nabla}\vec{\nabla} \cdot \vec{A} - (\xi - \varsigma)\frac{\hat{\varepsilon}_r^{-1}}{\varepsilon_0} : \vec{\nabla} \times \vec{A} \quad (7)$$

It is clear that for a medium with $\xi = \varsigma = 0$, which is a medium without ME effect, as well as a medium with $\xi = \varsigma \neq 0$ which is the case of an achiral but topological medium, the introduced generalized gauge theory is simplified to a normal Lorentz gauge. Using (7), (6) is further simplified to a Helmholtz equation in the form of:

$$\nabla^2 \vec{A} + k_0^2 \mu_r \hat{\varepsilon}_r : \vec{A} = 0 \quad (8)$$

in which $k_0 = \omega\sqrt{\varepsilon_0 \mu_0}$ is the wave vector of the light in free space. Solutions to (8) are called wave potentials. In a general anisotropic medium, it is not easy to solve (8) due to the coupling between the scalar components of the magnetic vector potential. In this paper I only discuss a biaxial anisotropic material, in which the optical axis is aligned with the coordinate system. For such a configuration the only non-zero components of the permittivity tensor are $\varepsilon_{r\,xx}$, $\varepsilon_{r\,yy}$ and $\varepsilon_{r\,zz}$.

In order to solve for eigenvalues of (8), the boundary conditions are to be satisfied, which means that we have to solve for the field components. Using (3), (5), and (7) we can write the electric field as:

$$\vec{E} = -i\omega\vec{A} + \frac{\hat{\varepsilon}_r^{-1}}{i\omega \varepsilon_0 \mu_0 \mu_r} : \vec{\nabla}\vec{\nabla} \cdot \vec{A} \\ -2i\gamma \frac{\hat{\varepsilon}_r^{-1}}{\varepsilon_0} : \vec{\nabla} \times \vec{A} \quad (9)$$

and the magnetic field $\vec{H}$ is derived using eqs. (2b), (3), and (4) as:

$$\vec{H} = \frac{1}{\mu_0 \mu_r}\vec{\nabla} \times \vec{A} - (\chi - i\gamma)\vec{E} \quad (10)$$

### 3. Optical waves at the surface of an anisotropic material with ME effect

We consider a semi-infinite problem in which the surface of an anisotropic, topological or chiral material lying at $z = 0$ supports 1-dimensional propagation of electromagnetic waves along the $x$-axis. The solutions for different components of the vector potential can be constructed as:

$$A_\alpha(\vec{r},\omega) = \tilde{A}_1^\alpha \exp\left(-\kappa_z^{(a)} z\right)\exp(-i\beta x) \quad (11a)$$

for $z \geq 0$, which is assumed to be air. Inside the material ($z \leq 0$) it is given by:

$$A_\alpha(\vec{r},\omega) = \tilde{A}_2^\alpha \exp\left(\kappa_z^{(\alpha,2)} z\right)\exp(-i\beta x) \quad (11b)$$

In which $\beta = \beta' - i\beta''$ is the complex propagation constant, $\left(\kappa_z^{(a)}\right)^2 = \beta^2 - k_0^2$, and $\left(\kappa_z^{(\alpha,2)}\right)^2 = \beta^2 - \varepsilon_{r\alpha\alpha}\mu_r k_0^2$. Here $\alpha \in (x, y, z)$. $\tilde{A}_1^\alpha$ and $\tilde{A}_2^\alpha$ are unknown coefficients for the vector potential expansions in each domain.

Computing the tangential field components using (9) and (10), the boundary conditions at the surface $z = 0$ can be satisfied. Obviously, only 4 equations exist for the tangential boundary conditions, while there exist 6 unknowns $\left(\tilde{A}_1^x, \tilde{A}_2^x\right)$, $\left(\tilde{A}_1^y, \tilde{A}_2^y\right)$, and also $\left(\tilde{A}_1^z, \tilde{A}_2^z\right)$. It is easily verified that with classifications of the modes as transverse magnetic along a certain direction, described precisely by $\text{TM}_x$ $\left(A_y = A_z = 0\right)$, $\text{TM}_y \left(A_x = A_z = 0\right)$, and $\text{TM}_z \left(A_x = A_y = 0\right)$ modes, the boundary conditions cannot be satisfied for each individual mode. The possibility of supporting the $\text{TM}_x$ mode at the interface is first examined in more details. Using (9) and (10) we obtain the boundary conditions for the $y$ – component of the electric field as $-2i\gamma \kappa_z^{(x,2)} \tilde{A}_2^x / \varepsilon_0 \varepsilon_{ryy} = 0$, since for $z \geq 0$ we have $\gamma = 0$. This leads to the only possibility $\tilde{A}_2^x = 0$ for guided modes. In other words, $\text{TM}_x$ grouping cannot lead to nonzero solutions. Considering also the $\text{TM}_y$ mode, the boundary condition for $x$ – component of the electric field dictates $-2\gamma\omega\mu_0 \kappa_z^{(y,2)} \tilde{A}_2^y / \varepsilon_{rxx} = 0$, which again leads to the only possibility of $\tilde{A}_2^y = 0$. Finally, for the case of $\text{TM}_z$ mode, we again consider the boundary condition for the $y$ – component of the electric field which leads to $2\gamma \beta \tilde{A}_2^z / \varepsilon_0 \varepsilon_{ryy} = 0$, or $\tilde{A}_2^z = 0$. So we need to hybridize the solutions in pairs; i.e. $(A_x, A_y)$, $(A_y, A_z)$, and $(A_x, A_z)$. Moreover, from these 3 groups we can further omit $(A_x, A_z)$, since this choice cannot also satisfy the boundary conditions. To understand this, it is sufficient to consider the boundary conditions for the $y$ – component of the electric field as well as the $x$-component of the magnetic field. For the former we obtain $\kappa_z^{(x,2)} \tilde{A}_2^x + i\beta \tilde{A}_2^z = 0$, while for the latter we have $\left(\kappa_z^{(x,2)}\right)^2 \tilde{A}_2^x + i\beta \kappa_z^{(z,2)} \tilde{A}_2^z = 0$, which can be only correct if we have $\kappa_z^{(x,2)} = \kappa_z^{(z,2)}$, which is not applicable to a general biaxial medium. So we are left with only two groups, which are named here $A^{xy}$ and $A^{yz}$ referring to the choice of the pairs $(A_x, A_y)$ and $(A_y, A_z)$, respectively.

For the case of $A^{xy}$ modes the boundary conditions for the tangential components of the field are obtained as:

$$\left(\kappa_z^{(a)}\right)^2 \tilde{A}_1^x = \frac{\left(\kappa_z^{(x,2)}\right)^2}{\varepsilon_{rxx}\mu_r} \tilde{A}_2^x - 2\omega\mu_0 \gamma \frac{\kappa_z^{(y,2)}}{\varepsilon_{rxx}} \tilde{A}_2^y \quad (12)$$

$$\tilde{A}_1^y = 2\gamma \frac{\kappa_z^{(x,2)}}{\omega\varepsilon_0\varepsilon_{ryy}} \tilde{A}_2^x + \tilde{A}_2^y \quad (13)$$

$$\kappa_z^{(a)}\tilde{A}_1^y = (\chi - i\gamma)\frac{\left(\kappa_z^{(x,2)}\right)^2}{i\omega\varepsilon_0\varepsilon_{rxx}\mu_r}\tilde{A}_2^x - \\ \left(\varepsilon_{rxx} + \eta_0^2\mu_r 2i\gamma(\chi - i\gamma)\right)\frac{\kappa_z^{(y,2)}}{\varepsilon_{rxx}\mu_r}\tilde{A}_2^y \tag{14}$$

$$-\kappa_z^{(a)}\tilde{A}_1^x = \left(\varepsilon_{ryy} + \eta_0^2\mu_r 2i\gamma(\chi - i\gamma)\right)\frac{\kappa_z^{(x,2)}}{\varepsilon_{ryy}\mu_r}\tilde{A}_2^x + \\ i\omega\mu_0(\chi - i\gamma)\tilde{A}_2^y \tag{15}$$

in which $\eta_0^2 = \mu_0/\varepsilon_0$ is the free space impedance. Consequently using (12) to (15), the characteristic equation for the propagation constant associated with $A^{xy}$ modes is obtained as:

$$\left(\kappa_z^{(a)} + \frac{\kappa_z^{(y,2)}}{\mu_r} + 2i\gamma(\chi - i\gamma)\eta_0^2\frac{\kappa_z^{(y,2)}}{\varepsilon_{rxx}}\right)\left(\frac{\varepsilon_{rxx}}{\mu_r} + \frac{\kappa_z^{(x,2)}}{\mu_r \kappa_z^{(a)}} + 2i\gamma(\chi - i\gamma)\eta_0^2\frac{\varepsilon_{rxx}}{\varepsilon_{ryy}}\right) \\ = -\eta_0^2\left((\chi - i\gamma)\frac{\kappa_z^{(x,2)}}{\mu_r} - 2i\gamma\kappa_z^{(a)}\frac{\varepsilon_{rxx}}{\varepsilon_{ryy}}\right)\left((\chi - i\gamma) - 2i\gamma\frac{\kappa_z^{(y,2)}}{\varepsilon_{rxx}\kappa_z^{(a)}}\right) \tag{16}$$

in which $\eta_0 = \sqrt{\mu_0/\varepsilon_0}$. From this equation the propagation constant of the $A^{xy}$ mode ($\beta^{xy}$) is obtained. For materials without ME effect, we have $\chi = \gamma = 0$. Inserting this into (16), we have two classes of modes with individual characteristic relations given by $\kappa_z^{(a)} + \kappa_z^{(y,2)}/\mu_r = 0$ and $\kappa_z^{(a)} + \kappa_z^{(x,2)}/\varepsilon_{rxx} = 0$, which result in effective refractive indices ($n_{\text{eff}} = \beta/k_0$) given by $n_{\text{eff}}^2 = \mu_r(\mu_r - \varepsilon_{ryy})/(\mu_r^2 - 1)$ and $n_{\text{eff}}^2 = \varepsilon_{rxx}(\varepsilon_{rxx} - \mu_r)/(\varepsilon_{rxx}^2 - 1)$, respectively. Obviously the latter is just the plasmon dispersion defined for ordinary rays in anisotropic materials. Especially for $\mu_r = 1$, this reduces to the simple SPP dispersion $n_{\text{eff}}^2 = \varepsilon_{rxx}/(\varepsilon_{rxx} + 1)$. The former relation defines the dispersion of a well-defined optical mode only if $\mu_r \neq 1$. It is easily seen that if $\varepsilon_{ryy} = 1$, this reduces to the simple dispersion for magnetic plasmons as $n_{\text{eff}}^2 = \mu_r/(\mu_r + 1)$, hence this relation defines the dispersion of magnetic plasmons for ordinary waves. This immediately provides us with the physics behind (16): The right term is only non-zero in the presence of the ME effect and results in the coupling of the magnetic and electric plasmon dispersions. Now we consider the $A^{yz}$ modes. The boundary conditions for the tangential components of the field are derived as:

$$i\beta\kappa_z^{(a)}\tilde{A}_1^z = -2\gamma\omega\mu_0\frac{\kappa_z^{(y,2)}}{\varepsilon_{rxx}}\tilde{A}_2^y - \frac{i\beta\kappa_z^{(z,2)}}{\varepsilon_{rxx}\mu_r}\tilde{A}_2^z \tag{17}$$

$$\tilde{A}_1^y = \tilde{A}_2^y + \frac{i\beta 2\gamma}{\omega\varepsilon_0\varepsilon_{ryy}}\tilde{A}_2^z \tag{18}$$

$$\kappa_z^{(a)} \tilde{A}_1^y = -\left(\varepsilon_{rxx} + \eta_0^2 \mu_r 2i\gamma(\chi - i\gamma)\right) \frac{\kappa_z^{(y,2)}}{\varepsilon_{rxx}\mu_r} \tilde{A}_2^y +$$
$$(\chi - i\gamma) \frac{\beta \kappa_z^{(z,2)}}{\omega \varepsilon_0 \varepsilon_{rxx}\mu_r} \tilde{A}_2^z \tag{19}$$

$$i\beta \tilde{A}_1^z = i\omega \mu_0 (\chi - i\gamma) \tilde{A}_2^y +$$
$$\left(\varepsilon_{ryy} + \eta_0^2 \mu_r 2i\gamma(\chi - i\gamma)\right) \frac{i\beta}{\varepsilon_{ryy}\mu_r} \tilde{A}_2^z \tag{20}$$

for the $y$ – component of the magnetic field.

Considering (17) to (20), the characteristic equation for the propagation constant associated with $A^{yz}$ modes is obtained as:

$$\left(\kappa_z^{(a)} + \frac{\kappa_z^{(y,2)}}{\mu_r} + 2i\gamma(\chi - i\gamma)\eta_0^2 \frac{\kappa_z^{(y,2)}}{\varepsilon_{rxx}}\right)\left(\frac{\varepsilon_{rxx}}{\mu_r} + \frac{\kappa_z^{(z,2)}}{\mu_r \kappa_z^{(a)}} + 2i\gamma(\chi - i\gamma)\eta_0^2 \frac{\varepsilon_{rxx}}{\varepsilon_{ryy}}\right)$$
$$= -\eta_0^2 \left((\chi - i\gamma)\frac{\kappa_z^{(z,2)}}{\mu_r} - 2i\gamma \kappa_z^{(a)} \frac{\varepsilon_{rxx}}{\varepsilon_{ryy}}\right)\left((\chi - i\gamma) - 2i\gamma \frac{\kappa_z^{(y,2)}}{\varepsilon_{rxx}\kappa_z^{(a)}}\right) \tag{21}$$

Note that the only difference between (16) and (21) is the replacement of $\kappa_z^{(x,2)}$ with $\kappa_z^{(z,2)}$ in the second term on the left side and also the first term on the right side. With the same approach as above, we can consider the limit when the material sustains no ME effect. At such surfaces, we again obtain the dispersion relation for a magnetic surface plasmon as $n_{\text{eff}}^2 = \mu_r(\mu_r - \varepsilon_{ryy})/(\mu_r^2 - 1)$, however the electric plasmon dispersion is replaced by $n_{\text{eff}}^2 = (\varepsilon_{rzz}\mu_r - \varepsilon_{rxx}^2)/(1 - \varepsilon_{rxx}^2)$ which is the plasmon dispersion for the extraordinary rays.

## 4. Optical waves in an anisotropic slab waveguide with ME effect

In order to investigate the propagation mechanisms of optical modes in a thin film within the domain $|z| \leq d$, the solutions for the vector potentials is constructed as:

$$A_\alpha(\vec{r},\omega) = \tilde{A}_3^\alpha e^{-\kappa_z^{(a)}(z-d)} e^{-i\beta x} \tag{22a}$$

for $z \geq d$, which is assumed to be air. Inside the material ($|z| \leq d$) it is given by:

$$A_\alpha(\vec{r},\omega) = \left(\tilde{A}_{2e}^\alpha \cos\left(\kappa_z^{(\alpha,2)}z\right) + \tilde{A}_{2o}^\alpha \sin\left(\kappa_z^{(\alpha,2)}z\right)\right) e^{-i\beta x} \tag{22b}$$

and for $z \leq -d$ we have

$$A_\alpha(\vec{r},\omega) = \tilde{A}_1^\alpha e^{+\kappa_z^{(a)}(z+d)} e^{-i\beta x} \tag{22c}$$

in which $\alpha \in (x, y, z)$, $\left(\kappa_z^{(a)}\right)^2 = \beta^2 - k_0^2$ and $\left(\kappa_z^{(\alpha,2)}\right)^2 = \varepsilon_{r\alpha}\mu_r k_0^2 - \beta^2$. The possible solutions are again constructed by assuming a pair of potentials as $(A_x, A_y)$ and $(A_y, A_z)$ referred to as $A^{xy}$ and $A^{yz}$, respectively. By satisfying the boundary conditions, the characteristic equation for the propagation constant $(\beta' - i\beta'')$ is computed. It becomes evident after some straight forward algebraic calculations that we can further decompose the $A^{xy}$ modes into two pairs with $(A_x^e, A_y^o)$ and $(A_x^o, A_y^e)$ where the choice of the unknown amplitudes for the vector potential in the region $|z| \leq d$ is represented by $(\tilde{A}_{2e}^x, \tilde{A}_{2o}^y)$ and $(\tilde{A}_{2o}^x, \tilde{A}_{2e}^y)$, respectively. We denote the former class as $A_{eo}^{xy}$ and the latter as $A_{oe}^{xy}$. The characteristic equation for the $A_{eo}^{xy}$ modes is obtained as :

$$-\eta_0^2 \left( \frac{-2i\gamma}{\varepsilon_{r\,yy}} \kappa_z^{(a)} + \frac{\chi - i\gamma}{\varepsilon_{r\,xx} \mu_r} \kappa_z^{(x,2)} \cot\left(\kappa_z^{(x,2)} d\right) \right) \times$$
$$\left( \chi - i\gamma + \frac{-2i\gamma}{\varepsilon_{r\,xx}} \frac{\kappa_z^{(y,2)}}{\kappa_z^{(a)}} \cot\left(\kappa_z^{(y,2)} d\right) \right) =$$
$$\left( \kappa_z^{(a)} + \left( \frac{1}{\mu_r} + \eta_0^2 \frac{2i\gamma(\chi - i\gamma)}{\varepsilon_{r\,xx}} \right) \kappa_z^{(y,2)} \cot\left(\kappa_z^{(y,2)} d\right) \right) \times$$
$$\left( \left( \frac{1}{\mu_r} + \eta_0^2 \frac{2i\gamma(\chi - i\gamma)}{\varepsilon_{r\,yy}} \right) + \frac{\kappa_z^{(x,2)}}{\varepsilon_{r\,xx} \mu_r \kappa_z^{(a)}} \cot\left(\kappa_z^{(x,2)} d\right) \right)$$
(23)

and for the $A_{oe}^{xy}$ modes:

$$-\eta_0^2 \left( \frac{2i\gamma}{\varepsilon_{r\,yy}} \kappa_z^{(a)} + \frac{\chi - i\gamma}{\varepsilon_{r\,xx} \mu_r} \kappa_z^{(x,2)} \tan\left(\kappa_z^{(x,2)} d\right) \right) \times$$
$$\left( \chi - i\gamma + \frac{2i\gamma}{\varepsilon_{r\,xx}} \frac{\kappa_z^{(y,2)}}{\kappa_z^{(a)}} \tan\left(\kappa_z^{(y,2)} d\right) \right) =$$
$$\left( \kappa_z^{(a)} - \left( \frac{1}{\mu_r} + \eta_0^2 \frac{2i\gamma(\chi - i\gamma)}{\varepsilon_{r\,xx}} \right) \kappa_z^{(y,2)} \tan\left(\kappa_z^{(y,2)} d\right) \right) \times$$
$$\left( \left( \frac{-1}{\mu_r} - \eta_0^2 \frac{2i\gamma(\chi - i\gamma)}{\varepsilon_{r\,yy}} \right) + \frac{\kappa_z^{(x,2)}}{\varepsilon_{r\,xx} \mu_r \kappa_z^{(a)}} \tan\left(\kappa_z^{(x,2)} d\right) \right)$$
(24)

We now consider the solutions for the $A^{yz}$ modes. Again it becomes evident that the solutions can be decomposed into two different symmetry groups, for which the choice of the unknown amplitudes for the vector potential in the region $|z| \leq d$ is represented by $(\tilde{A}_{2e}^y, \tilde{A}_{2e}^z)$ and $(\tilde{A}_{2o}^y, \tilde{A}_{2o}^z)$. These modes are denoted as $A_{ee}^{yz}$ and $A_{oo}^{yz}$, respectively. The characteristic equation for $A_{ee}^{yz}$ modes is obtained as:

$$-\eta_0^2\left(\frac{2i\gamma}{\varepsilon_{r\,yy}}\kappa_z^{(a)} + \frac{\chi - i\gamma}{\varepsilon_{r\,xx}\mu_r}\kappa_z^{(z,2)}\tan\left(\kappa_z^{(z,2)}d\right)\right)\times$$

$$\left(\chi - i\gamma + \frac{2i\gamma}{\varepsilon_{r\,xx}}\frac{\kappa_z^{(y,2)}}{\kappa_z^{(a)}}\tan\left(\kappa_z^{(y,2)}d\right)\right) =$$

$$\left(\kappa_z^{(a)} - \left(\frac{1}{\mu_r} + \eta_0^2\frac{2i\gamma(\chi - i\gamma)}{\varepsilon_{r\,xx}}\right)\kappa_z^{(y,2)}\tan\left(\kappa_z^{(y,2)}d\right)\right)\times$$

$$\left(\left(\frac{-1}{\mu_r} - \eta_0^2\frac{2i\gamma(\chi - i\gamma)}{\varepsilon_{r\,yy}}\right) + \frac{\kappa_z^{(z,2)}}{\varepsilon_{r\,xx}\mu_r\kappa_z^{(a)}}\tan\left(\kappa_z^{(z,2)}d\right)\right)$$

(25)

whereas for $A_{oo}^{yz}$ is obtained as:

$$-\eta_0^2\left(\frac{-2i\gamma}{\varepsilon_{r\,yy}}\kappa_z^{(a)} + \frac{\chi - i\gamma}{\varepsilon_{r\,xx}\mu_r}\kappa_z^{(z,2)}\cot\left(\kappa_z^{(z,2)}d\right)\right)\times$$

$$\left(\chi - i\gamma + \frac{-2i\gamma}{\varepsilon_{r\,xx}}\frac{\kappa_z^{(y,2)}}{\kappa_z^{(a)}}\cot\left(\kappa_z^{(y,2)}d\right)\right) =$$

$$\left(\kappa_z^{(a)} + \left(\frac{1}{\mu_r} + \eta_0^2\frac{2i\gamma(\chi - i\gamma)}{\varepsilon_{r\,xx}}\right)\kappa_z^{(y,2)}\cot\left(\kappa_z^{(y,2)}d\right)\right)\times$$

$$\left(\left(\frac{1}{\mu_r} + \eta_0^2\frac{2i\gamma(\chi - i\gamma)}{\varepsilon_{r\,yy}}\right) + \frac{\kappa_z^{(z,2)}}{\varepsilon_{r\,xx}\mu_r\kappa_z^{(a)}}\cot\left(\kappa_z^{(z,2)}d\right)\right)$$

(26)

In general, four symmetry groups for the optical modes in an anisotropic slab waveguide with ME effect are obtained. It is notable that these modes can be further decomposed into individual $TM_x$, $TM_y$, and $TM_z$ groups when the ME effect is absent; i.e. to set $\chi = \gamma = 0$ in (23) to (26). In this sense 6 individual groups of modes are distinguishable, with the characteristic equations for the $TM_x$ modes given by $\varepsilon_{r\,xx}\kappa_z^{(a)} = -\kappa_z^{(x,2)}\cot\left(\kappa_z^{(x,2)}d\right)$ and $\varepsilon_{r\,xx}\kappa_z^{(a)} = \kappa_z^{(x,2)}\tan\left(\kappa_z^{(x,2)}d\right)$ for the *even* and *odd* distributions of the $A_x$ vector potential component, whereas for the $TM_y$ modes we have $\mu_r\kappa_z^{(a)} = \kappa_z^{(y,2)}\tan\left(\kappa_z^{(y,2)}d\right)$ and $\mu_r\kappa_z^{(a)} = -\kappa_z^{(y,2)}\cot\left(\kappa_z^{(y,2)}d\right)$ for the *even* and *odd* distributions of the $A_y$ vector potential component, and finally for the $TM_z$ mode we have $\varepsilon_{r\,xx}\kappa_z^{(a)} = \kappa_z^{(z,2)}\tan\left(\kappa_z^{(z,2)}d\right)$ and $\varepsilon_{r\,xx}\kappa_z^{(a)} = -\kappa_z^{(z,2)}\cot\left(\kappa_z^{(z,2)}d\right)$ for the *even* and *odd* distributions of the $A_z$ vector potential component, respectively. That means that in the absence of the ME effect, each component of the magnetic vector potential can be used to satisfy all the boundary conditions, without the need to make pairs out of the components. Interestingly, when we have $\varepsilon_{r\,xx} = \varepsilon_{r\,zz}$, the $TM_x$ and $TM_z$ modes have similar characteristic equations, and so represent degenerate modes.

## 5. Results and discussions for optical modes at the surface

Equations (16) and (21) are to be solved to obtain the propagation constants $\beta^{xy}$ and $\beta^{yz}$ accordingly. The propagation constants are in general complex quantities $\beta^{\alpha\beta} = \beta'^{\alpha\beta} - i\beta''^{\alpha\beta}$, in which $\beta'^{\alpha\beta}$ is the phase constant, and $\beta''^{\alpha\beta}$ is the attenuation constant. As an example, the propagation constant of the optical modes at the surface of a material with $\varepsilon_{r\,yy} = \varepsilon_{r\,zz} = 3.42$, $\varepsilon_{r\,xx} = 1 - \left(\left(\omega/\omega_p - i\gamma^d/\omega_p\right)\omega/\omega_p\right)^{-1}$ is obtained, where $\gamma^d = 0.01\omega_p$, $\chi = 3\times 10^{-3}$ and $\gamma = 1\times 10^{-3}$. Such a material evidently supports a hyperbolic dispersion [31] as well as ME effect.

The computed dispersion diagram is shown in figure 1 (a). It is easily understood that only $A^{xy}$ modes are supported. Interestingly, two kinds of surface modes are supported by such a simple geometric waveguide, one showing a normal SPP dispersion proposed by a Drude metal, and the other demonstrating a photonic-like dispersion lying between the optical lines in free space and the material. In this figure, the depicted refractive index is $n_y = \sqrt{\varepsilon_{r\,yy}}$. The mode depicted as mode I has a field profile which is localized in one dimension (along the z-axis) at the apex, as it is expected from a normal SPP mode (see figure 1 (b)). Moreover, it is a hybrid mode for which the amplitude of both $A_x$ and $A_y$ components of the vector potential, and as a consequence all the electric field components are at the same range. However, the mode depicted as mode II has a strange field profile as shown in figure 1 (c). Especially $E_x$ demonstrates a minimum just few nanometers away from the surface, and again rises to its maximum values by penetrating into the material. It is also radiative inside the material, since its dispersion line is lying within the light cone specified by the optical axis $k_y = n_y \omega/c$. Figure 1 (d) shows the complete field profile for all the components of the electric field. The reason for such a field profile is understood by considering the analytical form of the electric field distribution. Especially $E_x$ inside the material $(z < 0)$ is given by:

$$E_x = \frac{-1}{i\omega\mu_0\mu_r\varepsilon_0\varepsilon_{r\,xx}} \left(\tilde{A}_2^x\left(\varepsilon_{r\,xx}k_0^2 - \beta^2\right)e^{\kappa_z^{(x,2)}z} + \tilde{A}_2^y 2\gamma\omega\mu_0\mu_r \kappa_z^{(y,2)} e^{\kappa_z^{(y,2)}z}\right)e^{-i\beta x} \quad (27)$$

which is composed of two terms, each representing a plane wave distribution. Whenever these two waves can interfere, which happens when they have similar magnitudes, they can form a field profile inside the material as shown in figure 1 (c). Moreover, the terms with the exponents $\kappa_z^{(y,2)}$ and $\kappa_z^{(x,2)}$ belong to the radiative and evanescent waves, respectively, which means that the interference only happens within a short distance from the surface. It is evident that when $\gamma = 0$ the electric field is composed of only one exponential term, and cannot show such a modulation in its intensity.

Despite these two modes, there exists also a continuum of radiative modes both in the air and material regions, as shown in figure 1 (a). Those modes also support interesting field profiles with a strong modulation in intensity within the material region. The spatial distributions of the different electric field components have been shown in figure 2, for two different radiative modes depicted as modes III and IV in figure 1 (a), at the normalized frequency $\omega/\omega_p = 1$. In a medium which is not chiral, the bulk modes represent normal plane wave patterns with a planar phase evolution. However, as it is shown in (9), the representative modes in a bulk chiral material cannot be expanded with simple linearly polarized plane waves, but rather a combination of two plane waves will be necessary. The superposition of these two waves can demonstrate an interference pattern with a non-planar distribution, as shown in figure 2.

## 6. Results and discussions for optical modes in slab wavguides

As an example system, the same parameters for the material as considered in the previous section are also assumed here, specifically we set $\varepsilon_{ryy} = \varepsilon_{rzz} = 3.42$, $\varepsilon_{rxx} = 1 - \left(\left(\omega/\omega_p - i\gamma^d/\omega_p\right)\omega/\omega_p\right)^{-1}$, where $\gamma^d = 0.01\omega_p$, $\chi = 3\times10^{-3}$ and $\gamma = 1\times10^{-3}$. The thickness of the slab waveguide is 400 nm, that is $d =$ 200 nm. In such a waveguide, all the modal groups, namely $A_{eo}^{xy}$, $A_{oe}^{xy}$, $A_{ee}^{yz}$, and $A_{oo}^{yz}$ are present. Figure 3 (a) shows the computed dispersion diagram for the $A_{eo}^{xy}$ group. The dispersion of this mode is quite similar to the dispersion of the guided plasmons in metallic thin films. However, there are two distinguished regions in the energy – momentum space: while the phase constant is asymptotic to the free–space optical line at lower frequencies, it crosses the optical line $k = n_y k_0$ at the normalized frequency of $\omega/\omega_p = 0.6$. In the region between these two optical lines, the field profile demonstrates the excitation of a photonic-like mode, for which the electric field is mostly concentrated inside the material, as shown in figure 3 (b). However, at the frequencies $\omega > 0.6\omega_p$, the optical modes behave more like the plasmonic modes, for which all the electric field components are localized (along the z-axis) at the surface. The computed dispersion diagram for the second class of the optical modes, namely $A_{oe}^{xy}$, is shown in figure 4 (a). Interestingly this modal class supports a complex mode diagram, apparent from the overlapping of the dispersion diagrams for the modes denoted by I and II, for the frequencies $\omega > 0.58\omega_p$, while at lower frequencies these two modes have different propagation constants. The reason behind this behaviour is the hybridization of the optical modes due to the ME effect. Without the ME effect, $TM_x$ and $TM_y$ modal classes are supported by the slab waveguide, where for the latter the only nonzero electric field component is $E_y$. So the $TM_y$ modes cannot provide the excitation of SPPs, but supports guided photonic modes because of the higher refractive index of the core in comparison with the surroundings. Unlike the $TM_y$ modes, $TM_x$ modes can support the excitation of SPPs. The interesting splitting in the dispersion diagram of figure 4 (a) happens because of the hybridization of these two modal classes induced by the ME effect. Moreover, this complex mode diagram happens only for the $A_{oe}^{xy}$ but not for $A_{eo}^{xy}$ modes. This is because that $E_y$ is the most dominant field component (see figures 4(b-d))), which is already localized at the interface because of its *odd* distribution for $A_{eo}^{xy}$ modes. This means that the deviation of the plasmonic and photonic field profiles in this case is not strong enough to cause the splitting of the dispersion diagram.

In contrast to the $A^{xy}$ modes, $A^{yz}$ modes do not support the excitation of SPPs, but rather only the photonic modes are excited, since $\varepsilon_{ryy}$ and $\varepsilon_{rzz}$ represent a dielectric – like permittivity. Figure 5 (a) shows the computed dispersion diagram for the $A_{ee}^{yz}$ modal classes. Another interesting complex mode diagram is visible, caused by the merging of the modes II and III for the energies $\omega > 0.66\omega_p$. Interestingly, both the splitting and the merging points happen when the dispersion diagram crosses the optical line in free space, which results in the splitting of the optical modes into a radiative and guided mode, for the normalized energies $0.54 < \omega/\omega_p < 0.66$. The field profiles for the depicted modes I to IV at various energies are shown in figures 5 (b-e). Clearly, the electric field component $E_y$ has *even* spatial distribution, but $E_x$ has an *odd* spatial distribution. As it is expected, the number of the field maxima is ascendingly ordered by the ordering of the modes. However, mode II demonstrates an exception for which the number of the maxima for the $E_x$ component is two, similar to mode I, but the $E_y$ profile demonstrates 3 maxima, as expected from mode III.

Finally, the computed dispersion diagram for $A_{oo}^{yz}$ is also shown in figure 6 (a). Only two guided modes for the considered frequency range are visible. All the mode profiles have *even* spatial distributions for the $E_x$ field component, but *odd* for the $E_y$ field component, in contrast to the $A_{ee}^{yz}$ modal class (see figure 6 (b) and (c)). Interestingly, while the mode I is a $TM_y$ - like mode for which the magnitude of the $E_y$ component is dominant, mode II is a hybrid mode, with similar range of amplitudes for all the electric field components.

## 7. Conclusion

In conclusion, the analytical treatment of the modal distributions at the interfaces and slabs composed of anisotropic materials with ME indices is reported by utilizing a vector potential approach and introducing a generalized Lorentz gauge. It is shown that 4 distinguished groups of the modes are supported by such waveguides, caused by the hybridization of the so-called transverse magnetic modes. The characteristic equations derived for computing the propagation constants are applicable to understanding the guided modes of the chiral materials like quartz, $AgGaS_2$, and $LiIO_3$, as well as investigating the axion electrodynamics in materials like $Bi_2Se_3$ and $BiTe_3$. Moreover, it has been shown that the ME effect can induce the formation of complex modes inside hyperbolic materials, and alteration of the usual modal distributions in thin films from photonic to plasmonic and vice-versa. In fact, engineering the scalar and pseudoscalar ME indices to reach this goal can open up new classes of metamaterials, which are both chiral in bulk and topological at the interfaces.

**Acknowledgment**
NT gratefully acknowledges the fruitful discussions with Prof. Dr. Peter van Aken (Max-Planck Institute for Solid State Research, Stuttgart, Germany) which enhanced greatly the clarity of the discussions in this work.

**References**
[1] Knight J C, Birks T A, Russell P S and Atkin D M 1996 All-silica single-mode optical fiber with photonic crystal cladding *Opt. Lett.* **21** 1547-9
[2] Vlasov Y A, O'Boyle M, Hamann H F and McNab S J 2005 Active control of slow light on a chip with photonic crystal waveguides *Nature* **438** 65-9
[3] Edwards B, Alu A, Young M E, Silveirinha M and Engheta N 2008 Experimental verification of epsilon-near-zero metamaterial coupling and energy squeezing using a microwave waveguide *Phys. Rev. Lett.* **100** 033903
[4] Silveirinha M G and Engheta N 2007 Theory of supercoupling, squeezing wave energy, and field confinement in narrow channels and tight bends using epsilon near-zero metamaterials *Phys. Rev. B* **76** 245109

[5] Talebi N and Shahabdi M 2008 Analysis of the propagation of light along an array of nanorods using the generalized multipole techniques *J. Comput. Theor. Nanos.* **5** 711-6
[6] Talebi N, Shahabadi M, Khunsin W and Vogelgesang R 2012 Plasmonic grating as a nonlinear converter-coupler *Opt. Express* **20** 1392-405
[7] Talebi N and Shahabadi M 2010 Spoof surface plasmons propagating along a periodically corrugated coaxial waveguide *J. Phys. D: Appl. Phys.* **43** 135302
[8] Talebi N, Shahabadi M and Hafner C 2006 Analysis of a lossy microring using the generalized multipole technique *Prog. Electromagn. Res.* **66** 287-99
[9] Yang Z Y, Xu Z N, Lu D S, Zhu D Q and Li P 2001 Simulation of optical waveguides with FDTD method *Fiber Optics and Optoelectronics for Network Applications* **4603** 42-7
[10] Koshiba M, Hayata K and Suzuki M 1987 Finite–element method analysis of microwave and optical waveguides—trends in countermeasures to spurious solutions *Electronics and Communications in Japan (Part II: Electronics)* **70** 96-108
[11] Hohenester U 2015 Quantum corrected model for plasmonic nanoparticles: A boundary element method implementation *Phys. Rev. B* **91** 205436
[12] Busch K, Konig M and Niegemann J 2011 Discontinuous Galerkin methods in nanophotonics *Laser Photon. Rev.* **5** 773-809
[13] Hu J and Menyuk C R 2009 Understanding leaky modes: slab waveguide revisited *Adv. Opt. Photonics* **1** 58-106
[14] Boriskina S V, Benson T M, Sewell P and Nosich A I 2002 Highly efficient full-vectorial integral equation solution for the bound, leaky, and complex modes of dielectric waveguides *IEEE J. Sel. Top. Quantum Electron.* **8** 1225-32
[15] Zhang J X, Zhang L D and Xu W 2012 Surface plasmon polaritons: physics and applications *J. Phys. D: Appl. Phys.* **45** 113001
[16] Dyakonov M I 1988 New type of electromagnetic wave propagating at an interface *Sov. Phys. JETP* **67** 714-6
[17] Harrington R F 2001 *Time-Harmonic Electromagnetic Fields*: (New York: Wiley-IEEE Press)
[18] Fiebig M 2005 Revival of the magnetoelectric effect *J. Phys. D: Appl. Phys.* **38** R123-R52
[19] Bassiri S, Papas C H and Engheta N 1988 Electromagnetic-wave propagation through a dielectric-chiral interface and through a chiral slab *J. Opt. Soc. Am. A* **5** 1450-9
[20] Robbie K, Brett M J and Lakhtakia A 1996 Chiral sculptured thin films *Nature* **384** 616-
[21] Karch A 2011 Surface plasmons and topological insulators *Phys. Rev. B* **83** 245432
[22] Tellegen B D H 1948 The gyrator, a new electric network element *Philips Res. Rep.* **3** 81-101
[23] Tellegen B D H 1948 The synthesis of passive, resistanceless 4-poles that may violate the reciprocity relation *Philips Res. Rep.* **3** 321-37
[24] Wilczek F 1987 Two applications of axion electrodynamics *Phys. Rev. Lett.* **58** 1799-802
[25] Esslinger M, Vogelgesang R, Talebi N, Khunsin W, Gehring P, de Zuani S, Gompf B and Kern K 2014 Tetradymites as natural hyperbolic materials for the near-infrared to visible *ACS Photonics* **1** 1285-9
[26] Essin A M, Moore J E and Vanderbilt D 2009 Magnetoelectric polarizability and axion electrodynamics in crystalline insulators *Phys. Rev. Lett.* **102** 146805
[27] Hehl F W, Obukhov Y N, Rivera J P and Schmid H 2008 Relativistic analysis of magnetoelectric crystals: Extracting a new 4-dimensional P odd and T odd pseudoscalar from Cr2O3 data *Phys. Lett. A* **372** 1141-6
[28] John L 1996 Optical properties of isotropic chiral media *Pure Appl. Opt.: Europ. Opt. Soc. P. A* **5** 417
[29] Soukoulis C M and Wegener M 2011 Past achievements and future challenges in the development of three-dimensional photonic metamaterials *Nat. Photonics* **5** 523-30
[30] Chang M-C and Yang M-F 2009 Optical signature of topological insulators *Phys. Rev. B* **80** 113304
[31] Cortes C L, Newman W, Molesky S and Jacob Z 2012 Quantum nanophotonics using hyperbolic metamaterials *J. Opt.* **14** 063001

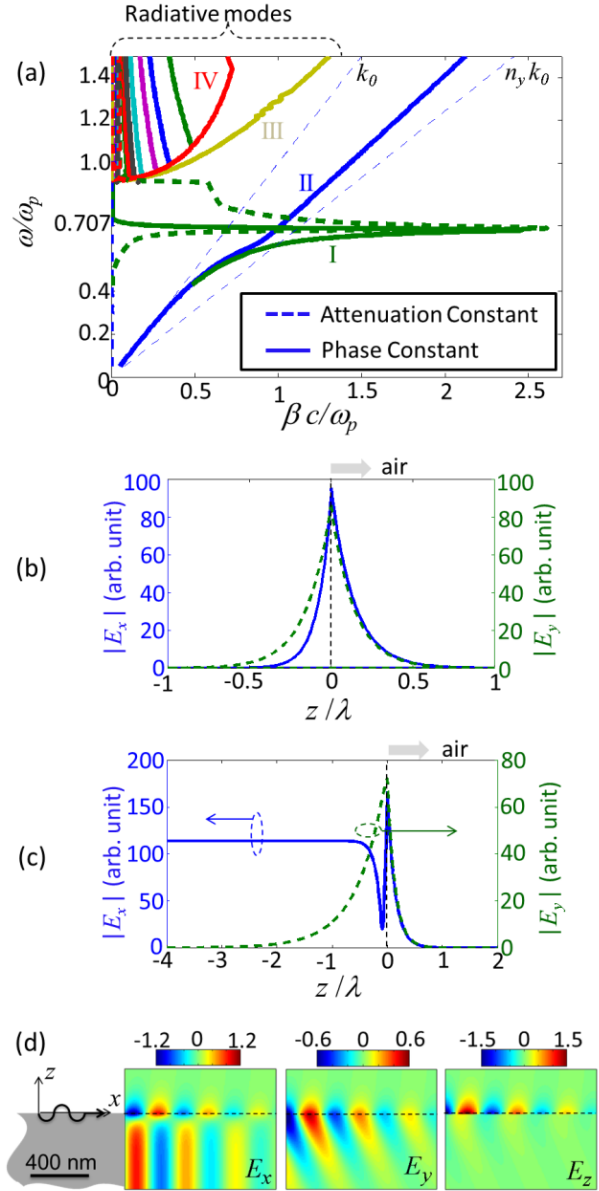

**Figure 1**. (a) The dispersion diagram of the optical modes supported at the interface of an anisotropic material with magneoelectric effect, with the parameters described in the text. The attenuation constant has been shown with thick dashed lines, while the thin dashed lines show the dispersion diagram. Magnitude of the $E_x$ and $E_y$ components of the electric field versus the $z$-axis, for (b) mode I and (c) modes II at the normalized frequency of $\omega/\omega_p = 0.6$. (d) Spatial distribution of the different electric field components for mode II at the normalized frequency of $\omega/\omega_p = 0.6$.

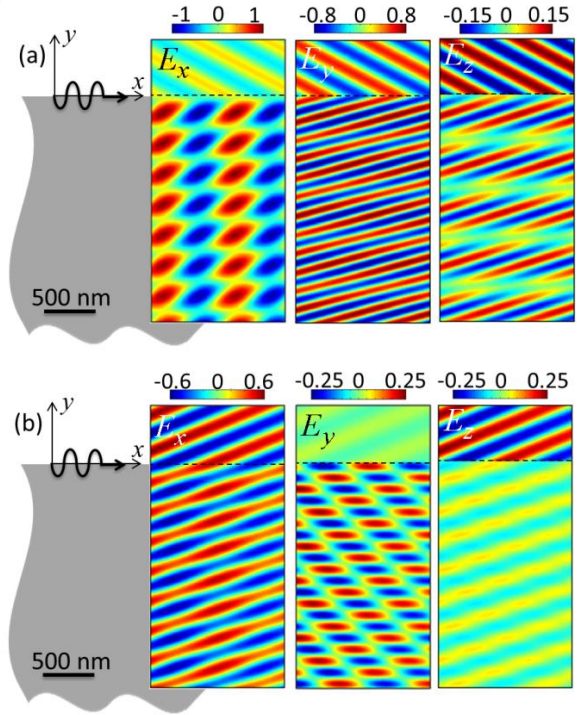

**Figure 2**. Spatial distribution of the electric field in the *xz*-plane, for the radiative modes depicted as (a) mode III and (b) mode IV in figure 1 (a), at the normalized frequency of $\omega/\omega_p = 1$.

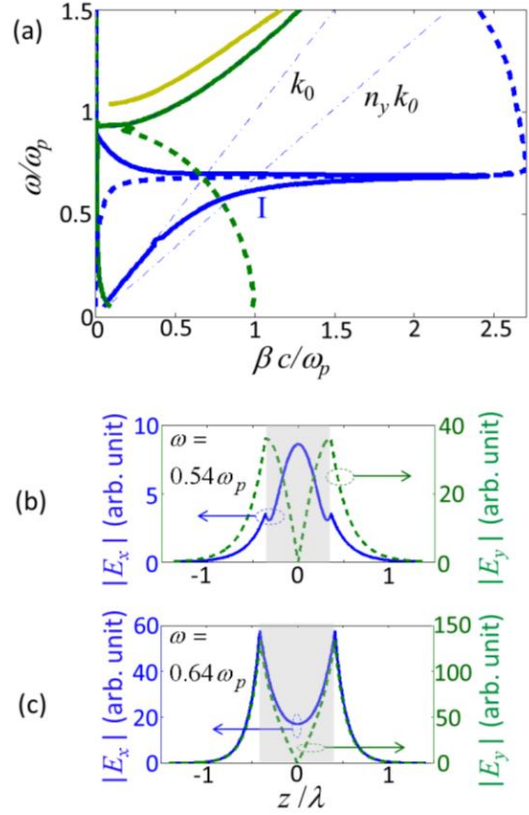

**Figure 3**. (a) Computed dispersion diagram of the $A_{xy}^{eo}$ modal group for an anisotropic slab waveguide with the parameters given in the text and the thickness of 400 nm. The thick solid and dashed lines show the phase constant and the attenuation constant respectively. Magnitude of the $E_x$ and $E_y$ components of the electric field versus $z$, for mode I at the normalized frequencies of (a) $\omega/\omega_p = 0.54$ and (c) $\omega/\omega_p = 0.64$.

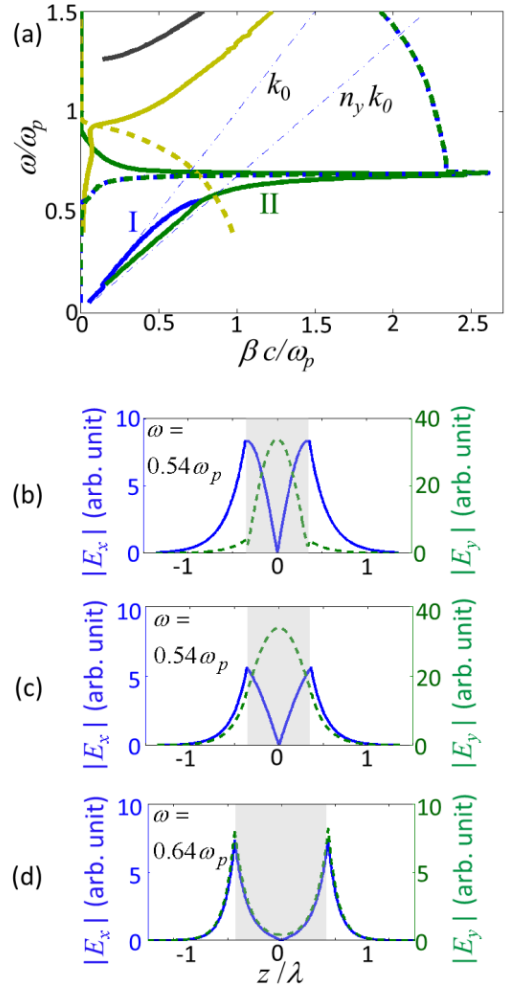

**Figure 4**. Computed dispersion diagram of the $A_{xy}^{oe}$ modal group for an anisotropic slab waveguide with the parameters given in the text and the thickness of 400 nm. The thick solid and dashed lines show the phase constant and the attenuation constant respectively. Magnitude of the $E_x$ and $E_y$ components of the electric field versus $z$, for (a) mode I and (b) mode II at the normalized frequency of $\omega/\omega_p = 0.54$, and (c) mode II at the normalized frequency of $\omega/\omega_p = 0.64$.

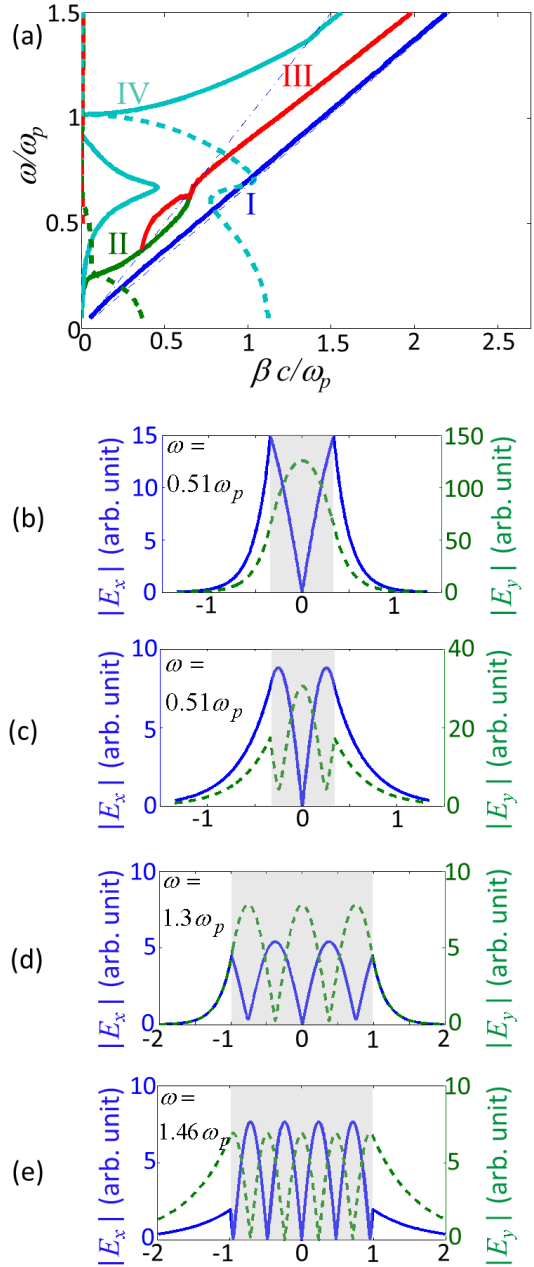

**Figure 5**. (a) Computed dispersion diagram of the $A_{yz}^{ee}$ modal group for an anisotropic slab waveguide with the parameters given in the text, and the thickness of 400 nm. The thick solid and dashed lines show the phase constant and the attenuation constant respectively. The thin dashed-doted lines show the optical lines in air and material. Magnitude of the $E_x$ and $E_y$ components of the electric field versus *z*, for (b) mode I, (c) mode II, (d) mode III, and (e) mode IV, at the frequencies depicted in the frames. The highlighted region shows the sab domain.

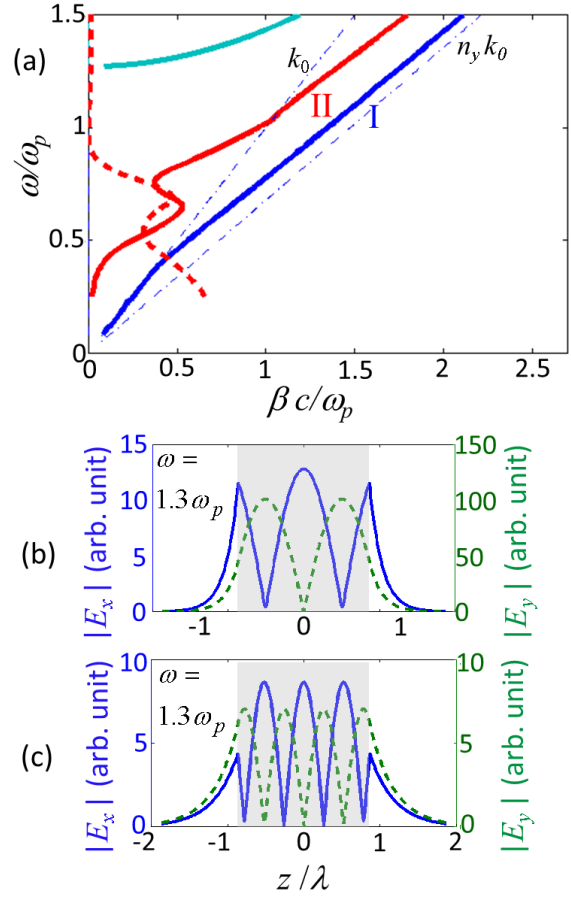

**Figure 6**. (a) Computed dispersion diagram of the $A_{yz}^{oo}$ modal group for an anisotropic slab waveguide with the parameters given in the text, and the thickness of 400 nm. The thick solid and dashed lines show the phase constant and the attenuation constant respectively. Magnitude of the $E_x$ and $E_y$ components of the electric field versus $z$, for (b) mode I and (c) mode II, at the frequencies depicted in the frames. The highlighted region shows the sab domain.